\documentclass[preprint,nofootinbib,showkeys,longbibliography]{revtex4-1}
\usepackage{epsfig}
\usepackage{amsmath}
\usepackage{amssymb}
\begin{document}
\author{Jacek Karwowski}
\affiliation {Institute of Physics, Faculty of Physics, Astronomy and
Informatics,\\
Nicolaus Copernicus University, Grudzi\c{a}dzka 5, 87-100 Toru\'n, Poland} 
\author{Andreas Savin}
\affiliation {Laboratoire de Chimie Théorique,
CNRS and Sorbonne University, 4 place Jussieu, 
75252 Paris cedex 05, France}
\title{Two-particle coalescence conditions revisited
\thanks{We dedicate this paper to Lutos\l{}aw Wolniewicz, 
an initiator of rigorous thinking in quantum chemistry.}}

\begin{abstract}
The notion of the $n$-th order local energy, generated by the $n$-th power of 
the Hamiltonian, has been introduced. The $n$-th order two-particle 
coalescence conditions have been derived from the requirements that the 
$n$-th order local energy at the coalescence point is non-singular and 
equal to the $n$-th power of the Hamiltonian eigenvalue. The first condition 
leads to energy-independent constraints. The second one is state-specific. 
The analysis has been done using a radial, one-dimensional, model 
Hamiltonian. The model is valid in the asymptotic region of $r\,\sim\,0$. 
The coalescence conditions set the relations between the expansion 
coefficients of the radial wave function into a power series with respect 
to $r$.\\
\end{abstract}


\maketitle

\section{Introduction}
\label{s1}

Two-particle Hamiltonians, since nearly a century, have been used as a 
playground for testing suitability of a variety of methods and models to 
the description of properties of simple quantum systems. The simplest ones, 
the hydrogen-like atom and the spherical harmonic oscillator not only 
served as a test of quantum mechanics, but also as a basis for the
development of analytical methods of solving the Hamiltonian eigenvalue 
problem and for the studies on the properties of its spectrum. By a proper 
change of coordinates, a two-particle Hamiltonian can be expressed as a sum 
relative motion. Consequently, the resulting two-particle eigenvalue problem 
can be separated to two independent one-particle problems: one describing the 
free motion of the centre of mass and the second one, describing the 
relative motion of the two particles. In general, an external potential 
prevents the separability. An exception is the parabolic confinement. 
Two-particle Schr\"odinger equations, independent of the form of the 
interaction potential, are separable also in parabolic external potentials. 
The interaction potential $V$ in the equation describing the relative motion 
depends only on the interparticle distance $r_{12}\,\equiv\,r$. Therefore, 
the Hamiltonian commutes with the angular momentum operators and, after the 
elimination of the angular part, its eigenvalue problem transforms to an 
infinite set of eigenvalue equations 
\begin{equation}
\label{01}
H\,\Phi_{\nu\lambda}(r)=E_{\nu\lambda}\,\Phi_{\nu\lambda}(r)
\end{equation}
of one-dimensional radial Hamiltonians
\begin{equation}
\label{02}
H=-\frac{1}{2\mu}\frac{d^2}{dr^2}+\frac{\lambda(\lambda+1)}{2\mu\,r^2}
+V(r),
\end{equation}
where $\lambda=0,1,2,\ldots$ is the angular momentum quantum number and
$\mu$ is the reduced mass.\footnote{It is convenient to use the radial 
Hamiltonian in the self-conjugate form which does not contain the 
first-order derivative.} In the case of two identical fermions the wave 
function is symmetric (a singlet pair) if $\lambda$ is even and
antisymmetric (triplet) if $\lambda$ is odd.

The point $r=0$ corresponds to the coalescence of the two particles. 
The information about the behaviour of the exact wave functions at 
this point is important both for the understanding of 
general properties of many-particle systems and for the construction of 
variational trial functions. Therefore, the subject attracted much interest 
- see, e.g. \cite{01,02,03,04,05,06,07,08,09,10,11,12,13,14} and references 
therein. Hamiltonian (\ref{02}) offers a simple and easy to treat
model. Though this model describes a "bare" pair of particles, under 
certain assumptions it can be generalized so that, after some 
modifications, it can be also applied to studies on the coalescence 
conditions in $N$-particle systems.  In particular, if the distance 
$r$ between two coalescing particles is much smaller than
the distance to any of the remaining $N-2$ particles, then one may 
expect that the influence of these particles on the properties of the 
coalescing pair can be described by a two-particle effective potential, 
parametrically dependent on the coordinates of the other particles.
After an approximate separation of the centre of mass of the two particles 
and a spherical averaging, we end up with a radial equation describing the 
pair of particles in the vicinity of the coalescence point.  From here one 
can derive constraints on the asymptotic form of the exact wave functions at 
the limit of $r=0$.  The best known of these constraints, Kato's cusp 
condition \cite{01}, can be derived from the requirement that in the case of 
two Coulomb-interacting particles the local energy is non-singular at $r=0$.  
Higher-order coalescence constraints have been obtained using some other 
universal properties of the exact wave functions in the vicinity of $r=0$ 
\cite{02,03,04,05}.  In a similar way the effects of the electron--electron 
coalescence on the electron density can be investigated.  The earliest works 
on this subject were published already half a century ago \cite{06}, but 
the links with the structure of the first-order density matrix and of the 
natural orbitals have been discovered very recently \cite{07,08,09}.
A detailed analysis of the wave function coalescence constraints, referred
to as {\em general coalescence conditions for the exact wave functions} has
been given by Kurokawa {\em et al.} \cite{10,11,12}.

A sensitive tool for the exploration of the behaviour of 
$\Phi_{\nu\lambda}(r)$ at $r=0$ is the local energy. Let
$\Psi_{\nu\lambda}(r)$ be a trial function which for specific values 
of parameters, and for $r<<1$, is equal to $\Phi_{\nu\lambda}(r)$. 
We define the $n$-th order local energy as
\begin{equation}
\label{03}
\frac{H^n\,\Psi_{\nu\lambda}(r)}{\Psi_{\nu\lambda}(r)}=
\mathcal{E}^{(n)}_{\nu\lambda}(r).
\;\;\;\;n=1,2,3,\ldots
\end{equation}
If $\Psi_{\nu\lambda}(r)=\Phi_{\nu\lambda}(r)$, i.e. it is the exact 
eigenfunction of $H$, then $\mathcal{E}^{(n)}_{\nu\lambda}(r)=E_{\nu\lambda}^n$.  
In this paper we derive the general two-particle coalescence conditions, as 
the ones of Kurokawa {\em et al.} \cite{10,11,12}, from the properties of 
the local energies at $r=0$.\footnote{Eq. (\ref{03}) is meaningful if 
$H^n\,\Psi(r)$ exists, i.e.  if $\Psi$ is $(2n)$-fold differentiable in its 
domain.  As shown by Fournais {\em et al.} \cite{15}, if the other electron 
coordinates do not coincide, then in a neighbourhood of the coalescence point 
Coulombic wave functions are analytic, i.e.  they are differentiable an 
arbitrary number of times.} We perform the analysis for a separable, model 
in which the radial part of the interaction is described by Hamiltonian 
(\ref{02}).  The constraints are derived using the information about the 
behaviour of the wave function at $r=0$. Therefore, the results are valid 
for both discrete and continuous spectra.  

For eigenfunctions of Hamiltonian (\ref{02}) the local energies of all orders 
have to be non-singular at the coalescence point. This property implies that 
the wave function has to compensate $r=0$ singularities generated by the 
Hamiltonian. The constraints imposed by the enforcement of this property are, 
for a given $\lambda$, energy-independent, i.e. they are common to all wave
functions $\Psi_{\nu\lambda}(r)$ which belong to the space spanned by the 
eigenfunctions of the radial Hamiltonian (\ref{02}). In the case of 
Coulomb-interacting particles and $n=1$ this constraint leads to Kato's 
cusp condition \cite{01}.

If at the coalescence point $\Psi_{\nu\lambda}(r)$ behaves as an 
eigenfunction of $H$ corresponding to the eigenvalue $E_{\nu\lambda}$ then
\begin{equation}
\label{04}
\left.\mathcal{E}_{\nu\lambda}^{(n)}(r)\right|_{r=0}=E_{\nu\lambda}^n. 
\end{equation}
This property is, by definition, energy-dependent. Therefore the constraints 
imposed by its enforcement are state-specific.

In the next section general coalescence conditions are derived and 
in Section~\ref{s3} an example of application is given. A graphical method of 
deriving explicit form of the energy-independent coalescence conditions is 
presented in the Appendix. Atomic units are used in this paper.

\section{Coalescence constraints}
\label{s2}

If we assume a Coulomb-like behaviour of $V(r)$ at $r=0$, expand it to a
power series about this point, and retain the first $q+2$ terms of the
expansion then we get 
\begin{equation}
\label{05}
V(r)=\sum_{p=-1}^q\,\alpha_p\,r^p,
\end{equation}
where, in the case of a parabolic confinement, $\alpha_2$ contains a 
contribution from the external potential. For Coulombic systems and for
$r<<1$, the term corresponding to $p=-1$ is dominant and determines the 
physical character of the potential.  If $\alpha_{-1}>0$ then the potential is 
repulsive (describing, for example, the interaction between two electrons); if 
$\alpha_{-1}<0$, it can describe an attractive electron--nucleus interaction.  
Formally, the potential parameters are unrestricted.  If $\alpha_{-1}=0$, 
then the potential loses its Coulombic character but, nevertheless, the 
higher-order coalescence conditions impose wave function constraints dependent 
on $\alpha_p$ with $p>0$ (cf. Subsection~\ref{s2.1.2}).~\footnote{See also an 
early study on the coalescence conditions for non-Coulombic potentials by Silanes 
{\em et al.} \cite{14}.} The higher-order terms in Eq. (\ref{05}) improve the 
analytic representation of the potential in the vicinity of the coalescence 
point, but have no physical meaning for large values of $r$.  For example, a 
term $\alpha_q\,r^q$, $\alpha_q>0$, $q\,\ge\,1$ prevents any kind of 
dissociation of the two particles.

For $r\,\sim\,0$ the radial functions describing the coalescing pair, can be 
expressed as
\begin{equation}
\label{06}
\Psi_{\nu\lambda}(r)\,\sim\,r^{\lambda+1}\,\psi_{\nu\lambda}(r),\;\;\;
\psi_{\nu\lambda}(0)\,\ne\,0,
\end{equation}
where $\nu$ refers to the energy eigenvalue and prefactor $r^{\lambda+1}$ 
compensates the singularity at $r\,\rightarrow\,0$ generated by the 
centrifugal potential $\lambda(\lambda+1)/r^2$ \cite{16}.
As the normalisation condition we set $\psi_{\nu\lambda}(0)=1$.
Since $\Psi_{\nu\lambda}(r)$ is analytic \cite{15} it can be expanded to
a power series of $r$.  We assume that $\psi_{\nu\lambda}(r)$ has an 
asymptotic expansion of order $s$ about $r=0$:
\begin{equation}
\label{07}
\psi_{\nu\lambda}(r)\,\thicksim\,\sum_{i=0}^s\,c_i\,r^i
\end{equation}
(to simplify notation, indices $\nu$ and $\lambda$ in $c_i$ are omitted). 
For $r<<1$, Eqs. (\ref{06}) and (\ref{07}), with properly defined 
expansion coefficients, give a correct representation of the 
eigenfunctions of $H$ at the vicinity of the coalescence point. Note that 
$\Psi_{\nu\lambda}(r)$ provides an asymptotic form of the corresponding 
eigenfunction but has no physical meaning outside of the coalescence region. 
First, the model is physically meaningful only for small $r$. Second, 
$\Psi_{\nu\lambda}(r)$, as defined in Eq. (\ref{06}), is not 
square-integrable in the range $r\in(0,\infty)$. 

Our aim is to derive the conditions limiting the values of the coefficients 
of expansion (\ref{07}) from two requirements defining the behaviour of
$\mathcal{E}_{\nu\lambda}^{(n)}(r)$ at $r=0$. The first group of these 
conditions, referred to as the {\em energy-independent constraints}, is a 
consequence of the requirement that $\mathcal{E}_{\nu\lambda}^{(n)}(r)$ 
is non-singular at $r=0$. The second group, the {\em energy-dependent 
constraints}, follows from the condition given by Eq. (\ref{04}).  
Since we are interested in the properties of the local energies at $r=0$, 
the behaviour of the wave functions outside of the asymptotic region 
is irrelevant for this discussion.

\subsection{Energy-independent constraints}
\label{s2.1}

We set
\begin{equation}
\label{08}
\Psi_{\nu\lambda}^{(0)}(r)=r^{\lambda+1}\,\psi_{\nu\lambda}^{(0)}(r),
\end{equation}
where 
\begin{equation}
\label{09}
\psi_{\nu\lambda}^{(0)}(r)\,=\,\sum_{i=0}^s\,c_i^{(0)}\,r^i,\;\;\;
c_i^{(0)}=c_i,
\end{equation}
and $c_i$ is defined in Eq. (\ref{07}). Hereafter, superscript $(0)$ in 
$c_i$ is usually dropped out. According to Eqs. (\ref{02}) and (\ref{05})
\begin{equation}
\label{10}
H\,r^{\lambda+1+i}=r^{\lambda+1}\,\left(A_i\,r^{i-2}+
\sum_{p=-1}^q\,\alpha_p\,r^{p+i}\right),
\end{equation}
where
\begin{equation}
\label{11}
A_i=-\frac{i(2\lambda+i+1)}{2\mu}.
\end{equation}
The combination of Eqs. (\ref{08}) and (\ref{10}) yields
\begin{equation}
\label{12}
H\Psi_{\nu\lambda}^{(0)}(r)=r^{\lambda+1}\left(\frac{c_{-1}^{(1)}}{r}+
\psi^{(1)}_{\nu\lambda}(r)\right),
\end{equation}
where
\begin{equation}
\label{13}
\psi^{(1)}_{\nu\lambda}(r)=\sum_{i=0}^{s+q}\,c^{(1)}_i\,r^i,
\end{equation}
\begin{eqnarray}
\label{14}
\nonumber
c_i^{(1)}&=&\sum_{p=-1}^q\,\alpha_p\,c_{i-p}^{(0)}+A_{i+2}\,c_{i+2}^{(0)},
\;\;\;i=-1,0,1,\ldots,s+q,\\
&&\mbox{with}\;\;c_k^{(0)}=0,\;\mbox{if}\;k<0,\;\mbox{or}\;k>s.
\end{eqnarray}
According to Eqs. (\ref{03}), (\ref{08}) and (\ref{12}).
\begin{equation}
\label{15}
\mathcal{E}_{\nu\lambda}^{(1)}(r)=\frac{H\Psi_{\nu\lambda}^{(0)}(r)}
{\Psi_{\nu\lambda}^{(0)}(r)}=\left(\frac{c_{-1}^{(1)}}{r}+
\psi^{(1)}_{\nu\lambda}(r)\right)\,\frac{1}{\psi^{(0)}_{\nu\lambda}(r)}.
\end{equation}
Since $\displaystyle \lim_{r\rightarrow{0}}\,\psi^{(0)}_{\nu\lambda}(r)=
c_0^{(0)}=1,\,$ and $\displaystyle \,\lim_{r\rightarrow{0}}\,\psi^{(1)}_
{\nu\lambda}(r)=c_0^{(1)}$,  the first-order local energy is asymptotically, 
at $r\,\sim\,0$, equal to
\begin{equation}
\label{16}
\mathcal{E_{\nu\lambda}}^{(1)}(r)
\genfrac{}{}{0pt}{1}{\thicksim}{r\rightarrow{0}}\,
\frac{c_{-1}^{(1)}}{r}+c^{(1)}_0,
\end{equation}
where, according to Eq. (\ref{14}), 
\begin{equation}
\label{17}
\begin{array}{lcl}
c^{(1)}_{-1}&=&\alpha_{-1}\,c^{(0)}_0+A_1\,c^{(0)}_1,\\
c_0^{(1)}&=&\alpha_{-1}\,c_1^{(0)}+\alpha_0\,c_0^{(0)}+A_2\,c_2^{(0)}.
\end{array}
\end{equation}
As it follows from Eq. (\ref{16}), $\mathcal{E}_{\nu\lambda}^{(1)}(0)$ is 
finite if $c^{(1)}_{-1}=0$. Therefore, the first order coalescence 
constraint reads
\begin{equation}
\label{18}
c_1^{(0)}=-\frac{\alpha_{-1}}{A_1}\,c_0^{(0)}.
\end{equation}
If $c^{(1)}_{-1}=0$ then
\begin{equation}
\label{19}
H\Psi_{\nu\lambda}^{(0)}(r)=r^{\lambda+1}\,\psi^{(1)}_{\nu\lambda}(r)\,
\equiv\,\Psi_{\nu\lambda}^{(1)}(r),
\end{equation}
and the procedure described by Eqs. (\ref{09}) -- (\ref{18}) may be
repeated with superscripts $(0)$  and $(1)$ replaced, respectively, by
$(1)$  and $(2)$. 

In general, if
\begin{equation}
\label{20}
c_{-1}^{(j)}=\alpha_{-1}\,c_0^{(j-1)}+A_1\,c_1^{(j-1)}=0,\;\;\mbox{for}
\;\;j=1,2,\ldots,n-1, 
\end{equation}
where
\begin{eqnarray}
\label{21}
c_i^{(j)}&=&\sum_{p=-1}^q\,\alpha_p\,c_{i-p}^{(j-1)}+
A_{i+2}\,c_{i+2}^{(j-1)},\;\;\;i=-1,0,1,\ldots,s+j\,q,\\
\nonumber
&&\mbox{with}\;\;c_k^{(j-1)}=0,\;\mbox{if}\;k<0,\;\mbox{or}\;k>s+(j-1)q,
\end{eqnarray}
then
\begin{equation}
\label{22}
H\Psi_{\nu\lambda}^{(j-1)}(r)=r^{\lambda+1}\,\psi^{(j)}_{\nu\lambda}(r)\,
\equiv\,\Psi_{\nu\lambda}^{(j)}(r),
\end{equation}
and
\begin{equation}
\label{23}
\mathcal{E}_{\nu\lambda}^{(n)}(r)\,\genfrac{}{}{0pt}{1}{\thicksim}
{r\rightarrow{0}}\,\frac{c_{-1}^{(n)}}{r}\,+\,c^{(n)}_0,
\end{equation}
where
\begin{equation}
\label{24}
\begin{array}{lcl}
c^{(n)}_{-1}&=&\alpha_{-1}\,c^{(n-1)}_0+A_1\,c^{(n-1)}_1,\\
c_0^{(n)}&=&\alpha_{-1}\,c_1^{(n-1)}+\alpha_0\,c_0^{(n-1)}+A_2\,c_2^{(n-1)}.
\end{array}
\end{equation}
From here we have\\

\noindent
\textbf{Theorem 1:} {\em The $n$-th order local energy
$\mathcal{E}_{\nu\lambda}^{(n)}(r)$ is finite at $r=0$ if} $c_{-1}^{(j)}=0$
for $j=0,1,2,\ldots,n$.\\
\textbf{Corollary:} {\em The coalescence constraints are independent of the
free parameter, $\alpha_0$, in the potential.}\\
Proof (by induction): $c_{-1}^{(1)}$ [Eq. (\ref{17})] does not depend on
$\alpha_0$. Assume, that $c_{-1}^{(j)}$, $j=1,2,\ldots,n-1$ do not depend on
$\alpha_0$. Then
\begin{eqnarray}
\label{25}
c_{-1}^{(n)}&=&\alpha_{-1}\,c_0^{(n-1)}+A_1\,c_1^{(n-1)}=
\alpha_0\left[\alpha_{-1}\,c_0^{(n-2)}+A_1\,c_1^{(n-2)}\right]+\\
&&\alpha_{-1}\left[\alpha_{-1}c_1^{(n-2)}+(A_1+A_2)c_2^{(n-2)}\right]+
A_1\left[\alpha_1\,c_0^{(n-2)}+A_3\,c_3^{(n-2)}\right].
\nonumber
\end{eqnarray}
According to Theorem 1, $\alpha_{-1}\,c_0^{(n-2)}+A_1\,c_1^{(n-2)}=
c_{-1}^{(n-1)}=0$. Therefore, $c_{-1}^{(n)}$ does not depend on $\alpha_0$.
$\blacksquare$.\\ 
Since $E_{\nu\lambda}$ can be included to $\alpha_0$, the Corollary implies
that the coalescence conditions derived from the requirement that the
$n$-th order local energies are non-singular at $r=0$ are, as 
expected, independent of the energy eigenvalues.

The coefficient $c_{-1}^{(1)}$ is a linear combination of $c_0^{(0)}$
and $c_1^{(0)}$ [Eq. (\ref{17})]. Similarly, $c_{-1}^{(2)}$ is a combination
of $c_0^{(1)}$ and $c_1^{(1)}$, i.e. of $c_j^{(0)}$, with $j=0,1,2,3$. As one 
can deduce from Eqs (\ref{21}), in order to obtain $c_{-1}^{(n)}$, 
one needs  $c_j^{(0)}$ with $j=0,1,\ldots,2n-1$. Therefore, the minimum 
value of $s$ in Eq. (\ref{07}) is $2n-1$. In practical terms, the upper 
limit for the highest order coalescence constraint is set by this 
condition - the coefficients of high powers of $r$ are ill defined in the 
asymptotic region because $r<<1$ implies that the high powers of $r$ are 
very small. Hereafter we set 
\[
s=2n-1.
\]

{\em Theorem 1} with Eqs. (\ref{20}) and (\ref{21}) yield explicit 
expressions for the energy-independent coalescence constraints:
\begin{equation}
\label{26}
c_{-1}^{(n)}=\sum_{i=0}^{2n-1}t_{ni}\,c_i=0,
\end{equation}
where\\[-20pt] 
\begin{eqnarray*}
&&t_{10}=\alpha_{-1},\;\;\;t_{11}=A_1,\\
&&t_{20}=\alpha_1\,A_1,\;\;\;t_{21}=\alpha_{-1}^2,\;\;\;
t_{22}=\alpha_{-1}\,(A_1+A_2),\;\;\;t_{23}=A_1A_3,\\
&&t_{30}=\alpha_{-1}^2\alpha_1+\alpha_{-1}\alpha_2(A_1+A_2)
+\alpha_3\,A_1A_3,\;\;\;
t_{31}=\alpha_{-1}\alpha_1(2\,A_1+A_2)+\alpha_2\,A_1A_3,\\
&&t_{32}=\alpha_{-1}^3+\alpha_1\,A_1(A_2+A_3),\;\;\;
t_{33}=\alpha_{-1}^2(A_1+A_2+A_3),\\
&&t_{34}=\alpha_{-1}(A_1A_3+A_1A_4+A_2A_4),\;\;\;t_{35}=A_1A_3A_5, 
\;\;\;\ldots.\\[-20pt]
\end{eqnarray*}
A graphical method of deriving $t_{mi}$ coefficients is presented in
the Appendix. From the first-order ($n=1$) constraint one can get the well 
known cusp condition generated by the Coulomb singularity:
\begin{equation}
\label{27}
c_1=-\frac{\alpha_{-1}}{A_1}\,c_0=\frac{\alpha_{-1}\,\mu}{\lambda+1}\,c_0.
\end{equation}
This condition defines the coefficient of the linear term of the expansion
of the radial function.  In the case of two-electron coalescence it is 
equivalent to Kato's cusp condition \cite{01}. For $n=2$ we get,
additionally, the second-order constraint:
\begin{equation}
\label{28}
-A_1\,A_3\,c_3=\alpha_1\,A_1\,c_0+\alpha_{-1}^2\,c_1+
\alpha_{-1}\left(A_1+A_2\right)\,c_2,
\end{equation}
where $c_1$ can be eliminated using Eq. (\ref{27}). 
In energy-independent constraints the {\em odd} coefficients, $c_1$, 
$c_3$, $c_5$, $\ldots$, can be expressed as linear functions of the 
{\em even} ones, $c_0$, $c_2$, $c_4$, $\ldots$:
\begin{eqnarray}
\nonumber
c_1&=&D_0^{(1)}c_0,\\
\label{29}
c_3&=&D_0^{(2)}c_0+D_2^{(2)}c_2,\\
\nonumber
c_5&=&D_0^{(3)}c_0+D_2^{(3)}c_2+D_4^{(3)}c_4.\\
\nonumber
\cdots& & \cdots \hspace*{5mm} \cdots \hspace*{5mm} \cdots
\end{eqnarray}
where 
\begin{equation}
\label{30}
D_0^{(1)}=-\frac{\alpha_{-1}}{A_1},\;\;\;\;\;\;
D_0^{(2)}=\frac{\alpha_{-1}^3}{A_1^2A_3}-\frac{\alpha_1}{A_3},
\;\;\;\;\;\;
D_2^{(2)}=-\frac{\alpha_{-1}(A_1+A_2)}{A_1A_3},\;\;\;\;\;\;
\ldots.
\end{equation} 
In general,
\begin{equation} 
\label{31} 
c_{2i+1}=\sum_{j=0}^{i}D^{(i+1)}_{2j}\,c_{2j},\;\;\;i=0,1,\ldots,n-1.
\end{equation} 
The asymptotic expansion of the wave function [Eq. (\ref{09})] becomes 
\begin{equation} 
\label{32} 
\psi^{(0)}_{\nu\lambda}(r)\,=\sum_{i=0}^{n-1}\,\left(c_{2i}+
r\,\sum_{j=0}^i\,D^{(i+1)}_{2j}\,c_{2j}\right)\,r^{2i},
\end{equation}
where we set $s=2n-1$. Alternatively, we can arrange the expansion according 
to the {\em even} coefficients:
\begin{equation} 
\label{33}
\psi^{(0)}_{\nu\lambda}(r)\,=
\,\sum_{i=0}^{n-1}\,c_{2i}\,W^{(2i)}(r)\,r^{2i},
\end{equation}
where
\begin{equation}
\label{34}
W^{(2i)}(r)=1+\sum_{j=i+1}^n\,D_{2i}^{(j)}\,r^{2j-1}.
\end{equation}

The local energy $\mathcal{E}^{(n)}$ does not diverge at $r=0$, regardless
of the values of $c_{2j}$, $j=0,1,2,\ldots,n-1$, if the {\em odd}  
expansion coefficients, $c_{2j+1}$, are defined as in Eq. (\ref{31}).  
As a consequence, the terms $c_{2i}\,r^{2i}$ in the 
expansion of the wave function are multiplied by polynomials $W^{(2i)}(r)$, 
containing only odd powers or $r$, with coefficients defined by the 
Hamiltonian parameters. 

\subsubsection{Pure Coulomb potential}
\label{s2.1.1}
In the pure Coulomb case, i.e. if $\alpha_p=0$ for $p\ne{-1}$, Eqs.
(\ref{21}) simplify to
\begin{equation}
\label{35}
c_i^{(j)}=\alpha_{-1}\,c_{i+1}^{(j-1)}+A_{i+2}\,c_{i+2}^{(j-1)}.
\end{equation}
Then,
\begin{eqnarray}
\nonumber
c_{-1}^{(n)}&=&\alpha_{-1}\,c_0^{(n-1)}+A_1\,c_1^{(n-1)}\\
\nonumber
&=&\alpha_{-1}\left[\alpha_{-1}\,c_1^{(n-2)}+A_2\,c_2^{(n-2)}\right]+
A_1\left[\alpha_{-1}\,c_2^{(n-2)}+A_3\,c_3^{(n-2)}\right]\\
\nonumber
&=&\alpha_{-1}^2\,c_1^{(n-2)}+\alpha_{-1}\left(A_1+A_2\right)c_2^{(n-2)}
+A_1A_3\,c_3^{(n-2)}\\
\nonumber
&=&\cdots \hspace*{5mm} \cdots \hspace*{5mm} \cdots\\
\label{36}
&=&\sum_{p=0}^j\,\alpha_{-1}^{n-j}\,\mathcal{A}_p^{(j)}\,c_{j+p-1}^{(n-j)}
=\cdots=\sum_{p=0}^n\,\alpha_{-1}^{n-p}\,\mathcal{A}_p^{(n)}\,c_{n+p-1},
\end{eqnarray}
where $\mathcal{A}_p^{(j)}$ is a $\binom{j}{p}$-term combination of  
$p$-fold products of $A_i$. In particular, $\mathcal{A}_0^{(n)}=1$, 
$\mathcal{A}_1^{(n)}=\sum_{i=1}^n\,A_i$,
$\;\mathcal{A}_n^{(n)}=\prod_{i=1}^n\,A_{2i-1}$,
$\;\mathcal{A}_2^{(3)}=A_1A_3+A_1A_4+A_2A_4$, 
$\mathcal{A}_2^{(4)}=\mathcal{A}_2^{(3)}+A_5(A_1+A_2+A_3)$,
$\mathcal{A}_3^{(4)}=A_1A_3A_5+A_1A_3A_6+A_1A_4A_6+A_2A_4A_6$, $\ldots$.
A graphical method of the evaluation of $\mathcal{A}_p^{(n)}$ is
given in the Appendix. 

For $n=4$ the constraints for the pure Coulomb potential read
\begin{eqnarray}
\nonumber
c^{(1)}_{-1}&=&\alpha_{-1}\,c_0+A_1\,c_1=0,\\[4pt]  
\nonumber
c^{(2)}_{-1}&=&\alpha_{-1}^2\,c_1+\alpha_{-1}(A_1+A_2)\,c_2+
A_1A_3\,c_3=0,\\[4pt]
\nonumber
c^{(3)}_{-1}&=&\alpha_{-1}^3\,c_2+\alpha_{-1}^2(A_1+A_2+A_3)\,c_3+
\alpha_{-1}(A_1A_3+A_2A_4+A_1A_4)\,c_4+\\[4pt]
\nonumber
&&A_1A_3A_5\,c_5=0,\\[4pt]
\label{37}
c^{(4)}_{-1}&=&\alpha_{-1}^4\,c_3+\alpha_{-1}^3(A_1+A_2+A_3+A_4)\,
c_4+\\[4pt]
\nonumber
&&\alpha_{-1}^2(A_1A_3+A_2A_4+A_1A_4+A_1A_5+A_3A_5+A_2A_5)\,c_5+\\[4pt]
\nonumber
&&\alpha_{-1}(A_1A_3A_5+A_1A_3A_6+A_1A_4A_6+A_2A_4A_6)\,
c_6+A_1A_3A_5A_7\,c_7=0.
\end{eqnarray}

\subsubsection{Potentials non-singular at $r=0$}
\label{s2.1.2}

Though potential (\ref{05}) with $\alpha_{-1}=0$ has no singularity at
$r=0$, it also generates singularities of higher-order local energies 
and coalescence constraints similar to the ones obtained for the singular
potential.  As it follows from Eq.  (\ref{24}), if $\alpha_{-1}=0$ 
then $c_1^{(n)}=0$. The remaining {\em odd} coefficients do not vanish.
According to Eq. (\ref{21}), $c_1^{(1)}=\alpha_1\,c_0+A_2\,c_3$.  As it 
was shown above, $c_1^{(1)}=0$.  Therefore, $A_3\,c_3=-\alpha_1\,c_0$, i.e. 
$c_3\,\ne\,0$ unless $\alpha_1=0$.  If $\alpha_{-1}=\alpha_1=0$, then 
$c_1=c_3=0$, but $A_5\,c_5=-\alpha_3\,c_0$, and so on. Using Eqs.  
(\ref{21}) one can prove by induction \\[8pt]
\textbf{Theorem 2:} {\em If $\alpha_{2p-1}=0$ for $p=0,1,2,\ldots,m$, 
then $c_{2p+1}^{(0)}=0$ for $p=0,1,2,\ldots,m$.}\\[8pt]
\textbf{Corollary:} If $V(r)$ contains only even powers of $r$, then
\begin{equation}
\label{38}
\psi^{(0)}(r)=\sum_{i=0}^{n-1}\,c_{2i}^{(0)}\,r^{2i}.
\end{equation}

\subsection{Energy-dependent coalescence constraints}
\label{s2.2}
 
The coalescence constraints considered so far depend on the angular 
momentum of the coalescing pair but do not depend on the energy 
eigenvalue. The constraints, expressed as linear relations between 
coefficients $c_i$ of the asymptotic expansions of the radial functions, 
are the same for all eigenfunctions of $H$. Therefore they have to be 
fulfilled also by the linear combinations of the eigenfunctions. 

We assume that the energy-independent constraints are fulfilled.
Consequently, the local energies of all orders from $1$ to $n$ are 
non-singular at $r=0$. The energy-dependent constraints result from 
the application of condition (\ref{04}). It is convenient to include 
$\alpha_0$ - the free parameter in the potential - to the eigenvalue.
We define
\begin{equation}
\label{39}
\tilde{H}=H-\alpha_0,\;\;\;
\epsilon=E_{\nu\lambda}-\alpha_0, 
\end{equation}
and
\begin{equation}
\label{40}
{\tilde{\mathcal{E}}}^{(n)}_{\nu\lambda}(r)=
\frac{\tilde{H}^n\,\Psi_{\nu\lambda}(r)}{\Psi_{\nu\lambda}(r)}=
\frac{(H-\alpha_0)^n\,\Psi_{\nu\lambda}(r)}{\Psi_{\nu\lambda}(r)}=
\sum_{j=0}^n\left(-1\right)^{n-j}\,\binom{n}{j}\,\alpha_0^{n-j}\,
\mathcal{E}^{(j)}_{\nu\lambda}(r),
\end{equation}
where, to simplify notation, indices $\nu$ and $\lambda$ 
in $\epsilon$ are omitted.\footnote{Note that a shift in the 
energy scale does not affect the eigenfunctions.} We assume that 
$\mathcal{E}^{(j)}_{\nu\lambda}(0)=E_{\nu\lambda}^j$ for $j=1,2,\ldots,n$. 
Then, according to Eqs. (\ref{04}) and (\ref{40}), 
\begin{equation}
\label{41}
\left.{\tilde{\mathcal{E}}}^{(n)}_{\nu\lambda}(r)\right|_{r=0}=
\left(E_{\nu\lambda}-\alpha_0\right)^n=\epsilon^n.
\end{equation} 
Therefore, formally, the replacement of $H$ by $\tilde{H}$ and 
$\mathcal{E}^{(n)}(0)$ by $\tilde{\mathcal{E}}^{(n)}(0)$ is equivalent
to setting $\alpha_0=0$ in Eq. (\ref{05}). Consequently, according to 
Eqs. (\ref{23}) and (\ref{24}), we can formulate\\[5pt]
\textbf{Theorem 3:}
{\em The $n$-th order energy-dependent constraints are expressed as}
\begin{equation}
\label{42}
\epsilon^j=c_0^{(j)}=
\alpha_{-1}\,c_1^{(j-1)}+A_2\,c_2^{(j-1)},\;\;\;j=1,2,\ldots,n.
\end{equation}

In particular,
\begin{eqnarray}
\label{43}
\epsilon&=&\alpha_{-1}\,c_1+A_2\,c_2=
-\frac{\alpha_{-1}^2}{A_1}\,c_0+A_2\,c_2,\\
\label{44} 
\epsilon^2&=&\alpha_{-1}\,c_1^{(1)}+A_2\,c_2^{(1)}\\
&=&\left(\alpha_{-1}\alpha_1+\alpha_2\,A_2\right)c_0+
\alpha_1\,A_2\,c_1+\alpha_{-1}^2\,c_2
\nonumber
+\alpha_{-1}\left(A_2+A_3\right)c_3+A_2\,A_4\,c_4,\\
\nonumber
\epsilon^3&=&\alpha_{-1}\,c_1^{(2)}+A_2\,c_2^{(2)}=
\sum_{i=0}^6f_i\,c_i,\\
\nonumber
\cdots&&\cdots\;\;\;\cdots,
\end{eqnarray}
where $f_i$ are linear combinations of products of the potential parameters
$\alpha_p$, $p=-1,1,2,3,4$ and $A_p$, $p=2,3,4,5,6$.
Using relation (\ref{43}), one can replace the eigenvalue parameter
$\epsilon$ in the energy-dependent constraints (\ref{42}) by $c_2$.

\subsection{The lowest-order constraints}
\label{s2.3}

For the reader's convenience, several first coefficients of the expansion 
of $\psi^{(0)}_{\nu\lambda}$ are given:
\begin{eqnarray}
\nonumber
c_0&=&1,\\
\nonumber
A_1\,c_1&=&-\alpha_{-1}\,c_0,\\
\nonumber
A_2\,c_2&=&\epsilon-\alpha_{-1}\,c_1,\\
\nonumber
A_1\,A_3\,c_3&=&-\alpha_1\,A_1\,c_0-\alpha_{-1}^2\,c_1-
\alpha_{-1}\left(A_1+A_2\right)\,c_2,\\
\nonumber
A_2\,A_4\,c_4&=&\epsilon^2-\left(\alpha_{-1}\alpha_1+\alpha_2\,A_2\right)
\,c_0-\alpha_1\,A_2\,c_1-\alpha_{-1}^2\,c_2-\alpha_{-1}\left(A_2+A_3\right)c_3,\\
\nonumber
A_1\,A_3\,A_5\,c_5&=&-\left[\alpha_{-1}^2\alpha_1+\alpha_{-1}\alpha_2
\left(A_1+A_2\right)+\alpha_3\,A_1\,A_3\right]\,c_0\\
\label{45}
&&-\left[\alpha_{-1}\alpha_1
\left(2\,A_1+A_2\right)+\alpha_2\,A_1\,A_3\right]\,c_1\\
\nonumber
&&-\left[\alpha_{-1}^3+\alpha_1\,A_1\left(A_2+A_3\right)\right]\,c_2
-\alpha_{-1}^2\left(A_1+A_2+A_3\right)\,c_3\\
\nonumber
&&-\alpha_{-1}\left(A_1\,A_3+A_1\,A_4+A_2\,A_4\right)c_4.
\end{eqnarray}

\section{Example}
\label{s3}

We consider two Coulomb-interacting particles in a parabolic confinement, 
i.e. we set $\alpha_{-1}\ne{0}$, $\alpha_2\ne{0}$ and $\alpha_i=0$ if 
$i\ne\,-1,\,2$. The radial Schr\"odinger equation (\ref{01}) reads
\begin{equation}
\label{46}
\left[-\frac{1}{2\mu}\frac{d^2}{dr^2}+\frac{\lambda(\lambda+1)}{2\mu\,r^2}
+\frac{\alpha_{-1}}{r}+\alpha_2\,r^2\right]\Phi_{\epsilon\lambda}(r)=
\epsilon\,\Phi_{\epsilon\lambda}(r),
\end{equation}
where subscript $\nu$  has been replaced by the corresponding energy
$\epsilon$. In the case of two electrons ($\mu=1/2$, $\alpha_{-1}=1$) the 
interaction is repulsive. In the case of two $\mu=1/2$ particles with 
opposite charges (electron--positron pair) $\alpha_{-1}=-1$ - the interaction 
is attractive. The spectrum of the confined system ($\alpha_2>0$) in both 
cases is purely discrete. In the unconfined systems ($\alpha_2=0$) the 
positive energy spectrum is continuous and the continuum 
spreads from $0$ to $\infty$. In the case of electron--positron pair discrete 
states with $\epsilon\,<\,0$ also appear. As it results from Eq. 
(\ref{46}), the transformation $r\,\rightarrow\,-r$  is equivalent to the 
replacement of $\alpha_{-1}$ by $-\alpha_{-1}$. Under this transformation
the wave function changes accordingly, but the eigenvalues remain the same. 
Note, that the last statement is valid only if the same eigenvalue 
{\em exists} in both repulsive and attractive case. In particular, if 
$\alpha_2=0$ then it is valid for continuous spectra. Otherwise, if 
$\alpha_2\,>\,{0}$, it is valid only for quasi-exact solutions of Eq. 
(\ref{46}) \cite{17}.

According to Eqs. (\ref{45}) the coefficients in the asymptotic expansion
(\ref{07}) are equal to
\begin{eqnarray}
\nonumber
A_1\,c_1&=&-\alpha_{-1},\\
\nonumber
A_1A_2\,c_2&=&\alpha_{-1}^2+\epsilon\,A_1,\\
\label{47}
A_1A_2A_3\,c_3&=&-\alpha_{-1}\left[\alpha_{-1}^2+\epsilon(A_1+A_2)\right],\\
\nonumber
A_1A_2A_3A_4\,c_4&=&\alpha_{-1}^2\left[\alpha_{-1}^2+
\epsilon(A_1+A_2+A_3)\right]+\epsilon^2\,A_1A_3-\alpha_2\,A_1A_2A_3,\\
\nonumber
A_1A_2A_3A_4A_5\,c_5&=&-\alpha_{-1}^3\left[\alpha_{-1}^2+
\epsilon(A_1+A_2+A_3+A_4)\right]\\
\nonumber
&&+\alpha_{-1}\,\epsilon^2\left(A_1A_3+A_2A_4+A_1A_4\right)
+\alpha_{-1}\,\alpha_2\,A_2A_3\left(A_1+A_4\right),\\
\nonumber
\cdots&&\cdots\;\;\;\cdots.
\end{eqnarray}
It is convenient to split expansion (\ref{07}) to two parts: the first 
one ($F_{\alpha_{-1}}$) one describing the interaction of unconfined 
particles and the second one ($\Delta_{\alpha_2}$), describing the effect 
of confinement: 
\begin{equation}
\label{48}
\psi_{\epsilon\lambda}(r)=F_{\alpha_{-1}}(r)+\Delta_{\alpha_2}(r)
\end{equation}
(in $F$ and $\Delta$ subscripts $\epsilon$ and $\lambda$ have been 
omitted). Using Eqs. (\ref{21}) and (\ref{45}) for $s=7$, $\alpha_{-1}=1$, 
and $\lambda=0$, i.e. for two electrons in a ${}^1S$ state, we get 
\begin{eqnarray}
\nonumber
F_1(r)&=&1+\frac{r}{2}+\left(\frac{1}{2}-
\epsilon\right)\,\frac{r^2}{6}
+\left(\frac{1}{8}-\epsilon\right)\,\frac{r^3}{18}
+\left[\left(\frac{1}{20}-\epsilon\right)\,\frac{1}{144}+
\frac{\epsilon^2}{120}\right]\,r^4\\
\nonumber
&+&\left[\left(\frac{1}{40}-\epsilon\right)\,
\frac{1}{2160}+\frac{23\,\epsilon^2}{10800}\right]\,r^5
+\left[\left(\frac{1}{70}-\epsilon\right)\,
\frac{1}{51840}+\frac{\epsilon^2}{720}\left(\frac{7}{45}-
\frac{\epsilon}{7}\right)\right]\,r^6\\
\label{49}
&+&\left[\left(\frac{1}{112}-\epsilon\right)\,
\frac{1}{1814400}+\frac{11\,\epsilon^2}{37800}\left(\frac{1}{24}-
\frac{\epsilon}{7}\right)\right]\,r^7+O\left(r^8\right),
\end{eqnarray}
and
\begin{equation}
\label{50}
\Delta_{\alpha_2}(r)=\frac{\alpha_2}{20}\,\left(r^4+
\frac{11}{30}\,r^5+\frac{61-130\,\epsilon}{1260}\,r^6
+\frac{59-498\,\epsilon}{17640}\,r^7\right)+
O\left(r^8\right).
\end{equation}
The expression for $\alpha_{-1}=-1$ can be obtained by the substitution 
$r\rightarrow-r$. The parabolic confinement does not affect $c_1$, 
$c_2$ and $c_3$. Therefore, up to the cubic term, the asymptotic expansion 
(\ref{07}) of $\psi_{\epsilon\lambda}$ is the same whether or not there is 
a parabolic confinement.
\begin{figure}
\includegraphics[width=0.49\textwidth]{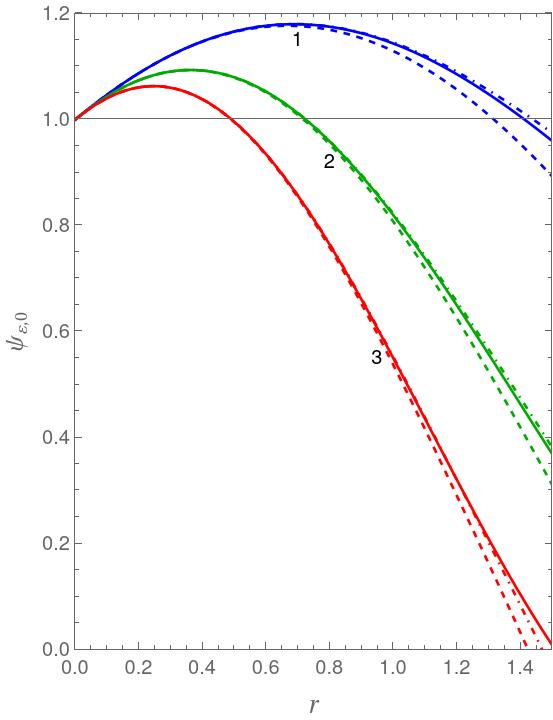}
\includegraphics[width=0.49\textwidth]{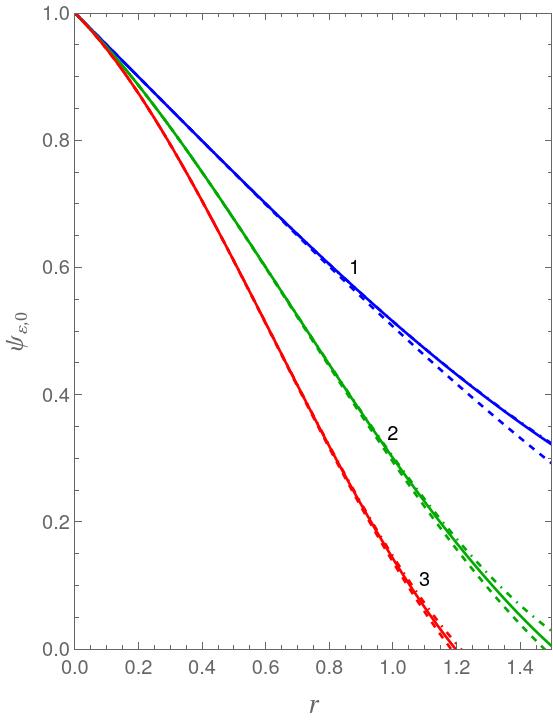}
\caption{\label{f0} 
Solid lines: the exact wave functions $\psi_{\epsilon,0}(r)$ of the first 
three ${}^1S$ states of harmonium (left panel) and of the confined positronium 
(right panel), versus $r$. Dashed lines - first eight expansion terms representing 
continuous spectrum wave functions of the unconfined systems [Eq. (\ref{49})]; 
dash-doted lines - the same, but including also the effect of confinement 
[Eq. (\ref{50})]. The confinement parameter $\alpha_2=1/4$; labels $1$, $2$, $3$, 
refer to the consecutive states.}
\end{figure}
\begin{figure}
\includegraphics[width=0.95\textwidth]{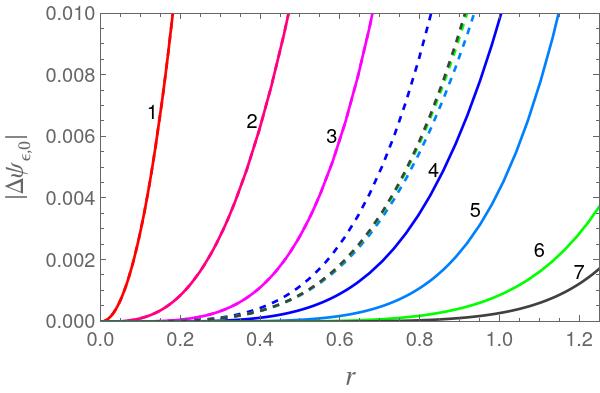}
\caption{\label{f0a} 
Differences between the exact ground state wave function 
of harmonium and the expansion given by Eq. (\ref{48}) for $\alpha_2=1/4$. 
Labels $1,2,\ldots,7$ are equal to $s$, the highest power of $r$ included in the
expansion. Results for the unconfined pair of electrons for $s=1,2,3$ are the
same as for the confined pair. The effect of ignoring confinement is indicated 
by dashed lines which correspond to $s=4,5,6,7$ and $\Delta_{\alpha_2}=0$.}
\end{figure} 

In Fig.~\ref{f0} the wave functions of the first three ${}^1S$ states of 
harmonium ($\epsilon=\,2230,\,4134,\,6074$ mH) and of the confined positronium 
($\epsilon=\,612,\,2805,\,4892$ mH) with $\alpha_2=1/4$, for $r<1.5$ bohr,
represented by solid lines, are compared with the results given by expansion 
(\ref{48}) including only the free-particle term $F_1$ (dashed lines), 
and also the confinement contribution  $\Delta_{\alpha_2}$ (dash-dotted lines). 
Surprisingly, the free-particle wave functions corresponding to the continuous 
spectra are nearly the same as the wave functions of the bound states of the 
confined systems, also for relatively large $r$. \footnote{Explicit 
expressions for the continuous spectrum wave functions can be found, e.g. in the 
monograph by Bethe and Salpeter \cite{18}. The expansion given by Eq.
(\ref{48}) with $\Delta_{\alpha_2}(r)=0$ is the same as the one obtained from 
the expansion of the exact eigenfunctions.} A contribution due to the confinement 
appears starting at $c_4$, but for both $c_4$ and $c_5$, it is an  
energy-independent constant term. 

The convergence pattern of expansion (\ref{48}) is shown in Fig.~\ref{f0a},
where differences between the exact ground state wave function of harmonium 
and the expansion (\ref{48}) with the number of terms varying from
$2$ to $8$ are plotted versus $r$. Line $1$ corresponds to the two-term
expansion, i.e. to Kato's cusp condition. Line $7$ corresponds to the 
$8$-term expansion including powers of $r$ from $0$ to $7$. The dashed 
lines refer to the unconfined pair of electrons (with term 
$\Delta_{\alpha_2}$  neglected).  

\section{Final remarks}
\label{s4}

We introduced the notion of the local energy of the $n$-th order (\ref{03}) 
and derived conditions which prevent the local energy of an arbitrary 
order to diverge at the coalescence point, referred to as the {\em
energy-independent coalescence constraints}. The wave function neither has to 
describe a bound state nor be a Hamiltonian eigenfunction. Only its asymptotic 
expansion at $r=0$ has to exist. By using the energy-independent coalescence 
constraints we can express the wave function in the vicinity of $
r=0$ as a linear combination of even powers of $r$ with each term of this 
combination modified by a polynomial composed of the odd powers with 
coefficients fixed by the coalescence constraints [Eqs. (\ref{32}), (\ref{33})]. 

From the requirement that the $n$-th order local energy at $r=0$
is proportional to the $n$th power of the eigenvalue, we derived the 
{\em energy-dependent constraints} fulfilled by the {\em even} coefficients 
of the expansion of the wave function. The coefficients $c_{2i}$,
$i=1,2,\ldots$, can be expressed as the $i$-th order polynomials of the
eigenvalue or, alternatively, of $c_2$. The complete set of the coalescence 
constraints is equivalent to the {\em general coalescence conditions} of 
Kurokawa {\em et al.} \cite{10,11,12}. 

In the coalescence region the Coulomb wave functions corresponding to the 
discrete spectrum of particles confined in a parabolic potential are nearly 
the same as the wave functions of the unconfined particles with the same 
energies, but belonging to the continuous part of the spectrum. This effect 
depends on the strength of confinement. The wave function of harmonium behaves 
as the wave function of two electrons for small $r$ and as the wave function of    
the harmonic oscillator for large $r$. The range of $r$ where the harmonic 
oscillator behaviour dominates extends with increasing $\alpha_2$.
For a moderate confinement (as e.g. $\alpha_2=1/4$), differences 
between the wave functions with confinement effects included and neglected  
only become noticeable for $r>1$.

\section*{Acknowledgement}
\noindent
We thank Dr.  Heinz-Jürgen Flad (Technische Universität München) for useful
discussions. 

\section*{Disclosure statement}
\noindent
No potential conflict of interest was reported by the authors.

\section*{ORCID}
\noindent
Jacek Karwowski: https://orcid.org/0000-0003-1508-2929\\
Andreas Savin: https://orcid.org/0000-0001-8401-8037

\bibliography{mrev}

\appendix

\section{Graphical representations}
\label{a}

\begin{figure}[b]
\includegraphics[width=0.99\textwidth]{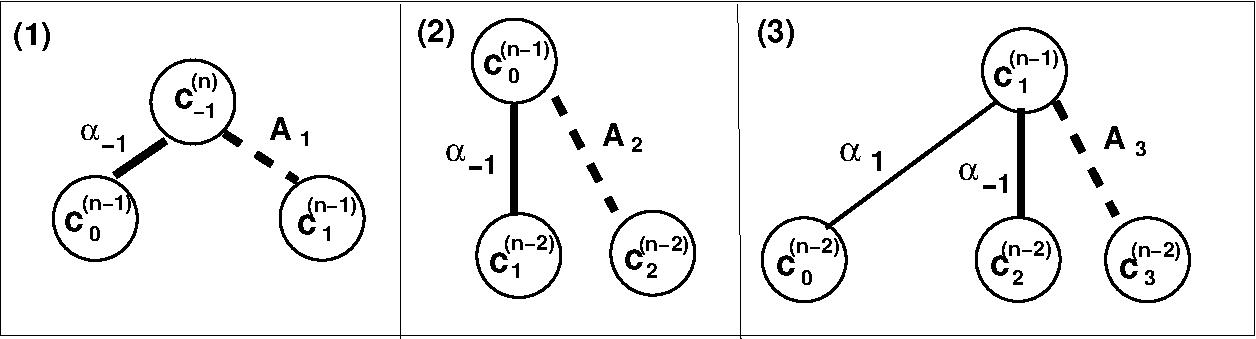}
\caption{\label{f1} 
Graphical representations of Eqs. (\ref{a1}) - panel (1), (\ref{a2}) - panel
(2), and (\ref{a3}) - panel (3). Arcs corresponding to $\alpha_0$ are not 
shown since they do not contribute to the constraints.}
\end{figure}

\begin{figure}
\includegraphics[width=0.8\textwidth]{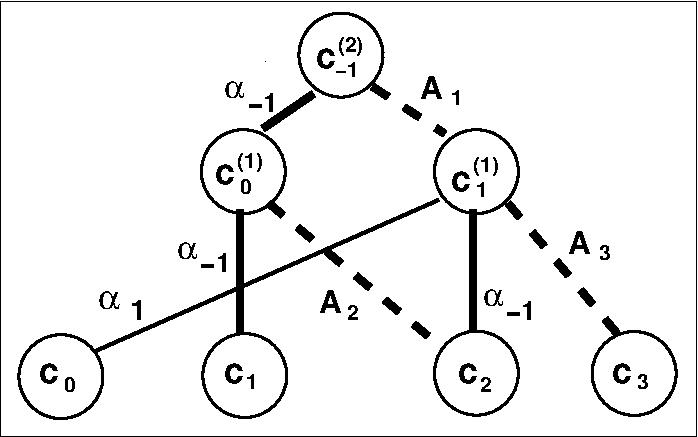}
\caption{\label{f2} 
Graphical representation of Eq. (\ref{a4}).} 
\end{figure}

\begin{figure}
\includegraphics[width=1.0\textwidth]{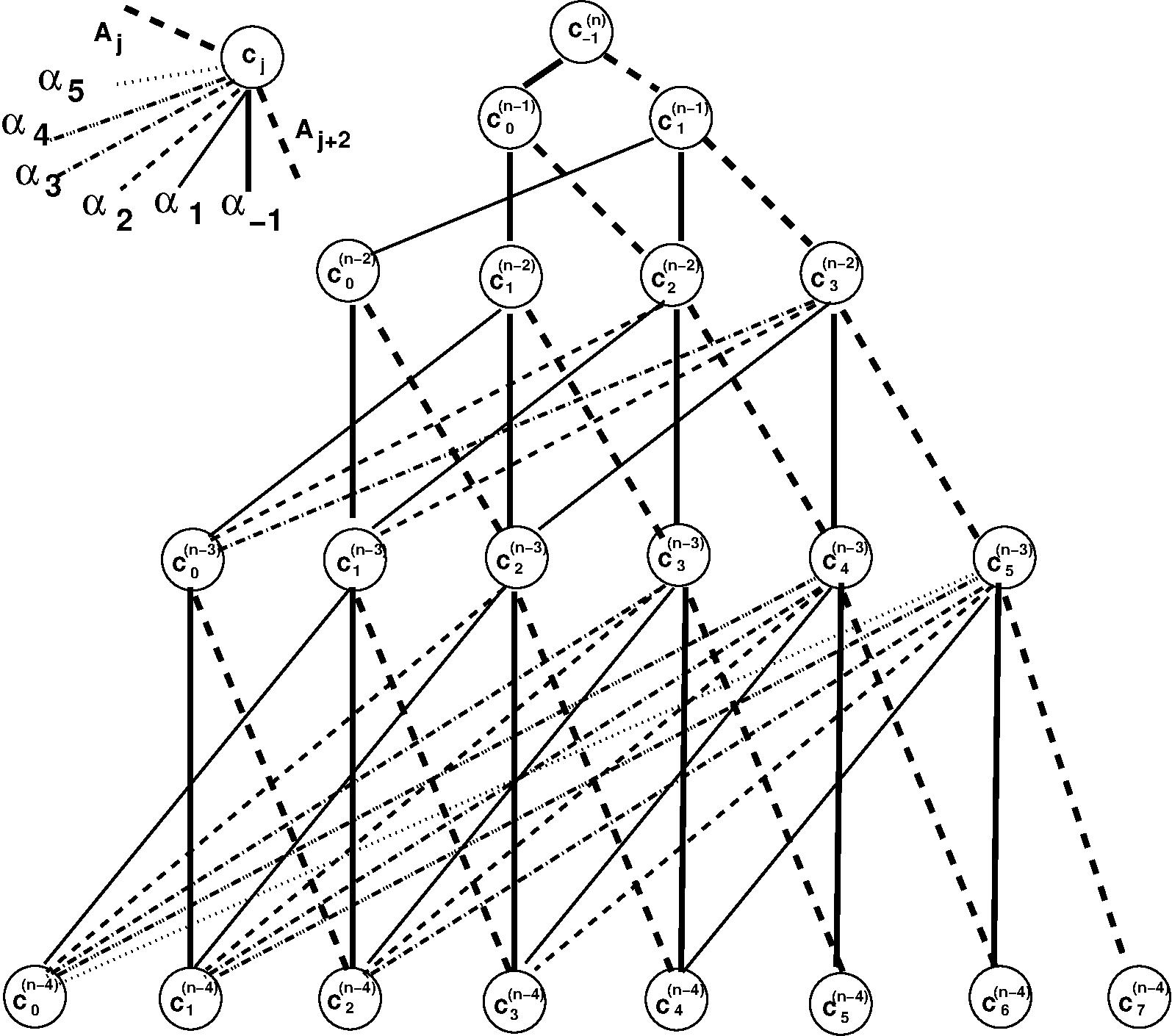}
\caption{\label{f3}
The first five rows of graphical representation of Eq. (\ref{26}). 
Arcs corresponding to $\alpha_0$ are not shown since they do not contribute 
to the constraints (see Theorem 1).}
\end{figure}

\begin{figure}[t]
\begin{center}
\includegraphics[width=0.75\textwidth]{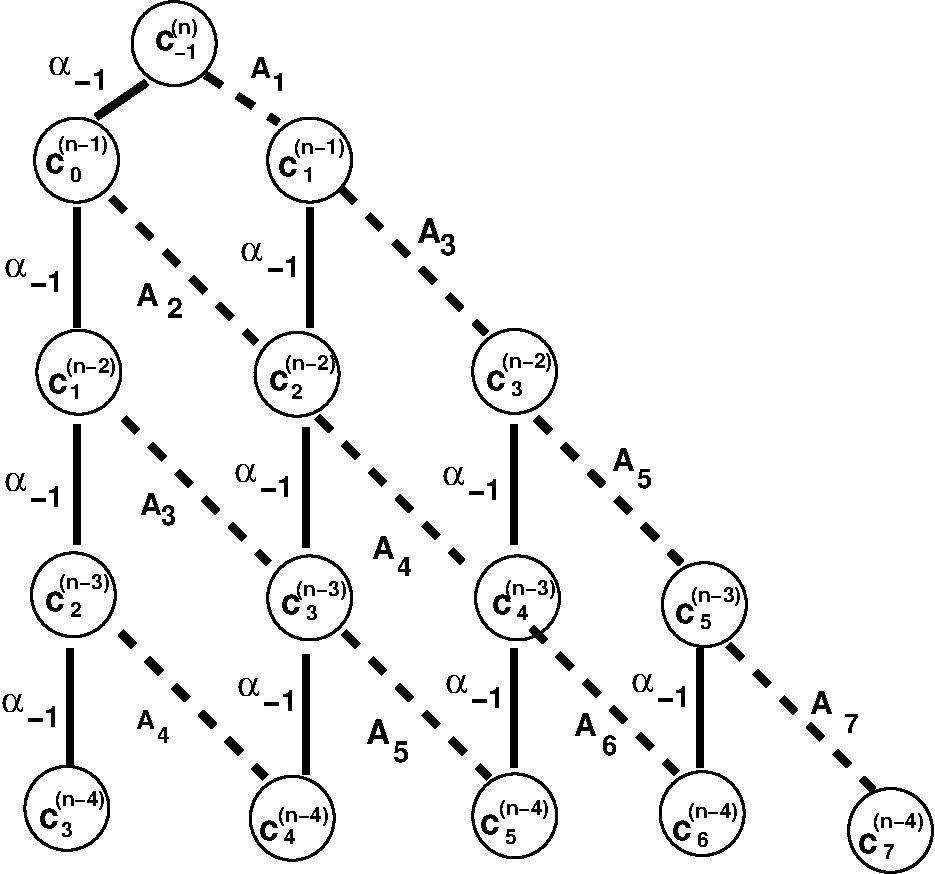}
\end{center}
\caption{\label{f4} Graph for the case of the pure Coulomb potential 
($\alpha_p=0$ if $p\ne{-1}$).}
\end{figure}

The derivation of energy-independent coalescence conditions is facilitated by a
graphical representation of Eqs. (\ref{21}) and (\ref{26}). The graphs are 
composed of {\em vertices} distributed in rows and linked by {\em arcs}. 
A vertex corresponds to a specific coefficient $c_i^{(j)}$, shown in the vertex. 
To each arc we assign an {\em arc index}. Sub-graphs composed of vertex $c_i^{(n)}$
and vertices $c_{j}^{(n-1)}$ together with arcs linking these vertices are
referred to as {\em segments}. Segments representing equations
\begin{eqnarray}
\label{a1}
c_{-1}^{(n)}&=&\alpha_{-1}\,c_0^{(n-1)}+A_1\,c_1^{(n-1)},\\
\label{a2}
c_0^{(n-1)}&=&\alpha_{-1}\,c_1^{(n-2)}+A_2\,c_2^{(n-2)},\\ 
\label{a3}
c_1^{(n-1)}&=&\alpha_1\,c_0^{(n-2)}+\alpha_{-1}\,c_2^{(n-2)}+A_3\,c_3^{(n-2)}, 
\end{eqnarray}
i.e. Eqs. (\ref{21}) for $i=-1,0,1$, are shown, respectively,
in panels (1), (2), (3) of Fig.~\ref{f1}. 
Equation
\begin{equation}
\label{a4}
c_{-1}^{(2)}=\alpha_{-1}\,c_0^{(1)}+A_1\,c_1^{(1)}=
\alpha_1\,A_1\,c_0+\alpha_{-1}^2\,c_1+
\alpha_{-1}\,\left(A_1+A_2\right)\,c_2+A_1\,A_3\,c_3.
\end{equation}
is represented in Fig.~\ref{f2} - the graph has been obtained by connecting
three segments of Fig.~\ref{f1} (for $n=2$) into one diagram. 

In general, expression (\ref{26}) for $c^{(n)}_{-1}$ is equal to the sum of
products of the arc indices and the coefficients $c_i^{(n)}$, taken over all
paths leading from $c^{(n)}_{-1}$ to all vertices of the selected level in
the graph (in one level there are vertices corresponding to a given order of
the local energy). Vertices of adjacent levels are linked by arcs - if
$c_a^{(j)}=\cdots+h\,c_b^{(j-1)}+\cdots$, then vertices $c_a^{(j)}$ and
$c_b^{(j-1)}$ are linked by an arc and the arc index is equal $h$. By the
construction, only paths going down from the uppermost vertex are allowed.

The uppermost part (the first five rows) of the most general graph (all
$\alpha_p\,\ne\,0$) is shown in Fig~\ref{f3}. Contributions from
$\alpha_{-1}$ and from $A_j$ are present in all orders. Contributions from
$\alpha_1$ start from the second order. From the third order up, we have
also contributions from $\alpha_2$ and $\alpha_3$. In the next order
contributions from $\alpha_4$ and $\alpha_5$ appear. And so on - each next
order activates two more terms of the expansion of $V(r)$.

The graph corresponding to the pure Coulomb potential, i.e. to the case of
$\alpha_p=0$ if $p\,\ne\,-1$, is given in Fig.~\ref{f4}. It is isomorphic
with the Pascal triangle. There are $\binom{j}{p}$ paths one can reach node
$c_{j+p-1}^{(n-j)}$ starting from node $c_{-1}^{(n)}$. With each path we
associate a product of all arc indices $A_i$ taken along this path. The
coefficient $\mathcal{A}_p^{(j)}$ introduced in Eq. (\ref{36}) is equal to the
sum of these products extended over all $\binom{j}{p}$ paths. For example,
nodes $c_{-1}^{(n)}$ and $c_6^{(n-4)}$ are linked by $\binom{4}{3}=4$ paths
and $\mathcal{A}^{(j)}_p=A_1A_3A_5+A_1A_3A_6+A_1A_4A_6+A_2A_4A_6$.
\end{document}